\documentclass{eas}
\usepackage{graphicx}
\begin{document}

\title{Stellar polarimetry with ESPaDOnS} 
\runningtitle{Petit \etal: Stellar polarimetry with ESPaDOnS} 
\author{P. Petit}
\address{Observatoire Midi-Pyr\'en\'ees, 14 avenue Edouard Belin, 31400 Toulouse, France \& Centro de Astrofisica da Universidade do Porto, rua das Estrelas, 4150-762 Porto, Portugal ({\tt petit@astro.up.pt})}
\author{J.-F. Donati}
\address{Observatoire Midi-Pyr\'en\'ees, 14 avenue Edouard Belin, 31400 Toulouse, France ({\tt donati@ast.obs-mip.fr})}
\author{the ESPaDOnS project team}
\address{Observatoire Midi-Pyr\'en\'ees and Observatoire de Paris-Meudon (France), University of Western Ontario (Canada), ESA (Netherlands)}
\begin{abstract}

ESPaDOnS is a new-generation cross-dispersed \'echelle spectropolarimeter, the commissioning phase of which is scheduled at the Canada-France-Hawaii Telescope (CFHT) for spring 2004.  This instrument will provide full coverage of the optical domain (370~nm to 1,000~nm) in all polarization states (circular and linear) at a resolving power of about 70,000, with a peak efficiency of 20\% (telescope and detector included).  It includes a bench-mounted spectrograph, fiber-fed from a Cassegrain-mounted module including all polarimetric and calibration facilities.

ESPaDOnS should be the most powerful tool dedicated to stellar spectropolarimetry, therefore opening unprecedented perspectives for major issues of stellar physics, from studies of stellar interiors to investigations of stellar atmospheres, stellar surfaces, stellar magnetic fields, and to observations of circumstellar environments and extra-solar planets.

\end{abstract}
\maketitle

\section{Introduction}

The combination of a polarimeter and a high-resolution \'echelle spectrograph constitutes a powerful tool for investigating the physics of stellar atmospheres and environments. In particular, stellar magnetic fields are mostly detected by their polarizing effect on spectral lines. The analysis of this so-called ``Zeeman effect'' requires polarized, high resolution spectra, so that polarization across spectral line profiles can be investigated in detail, in order to measure both field strength and orientation. A spectral resolution in excess of 50,000 (to obtain at least two resolved elements in the thermal width of a typical spectral line) and the possibility to record linearly and circularly polarized spectra are therefore requested for optimal investigations of stellar magnetism. In addition to high resolution, a wide spectral coverage (preferentially covering the whole optical domain) is needed by spectroscopic studies (polarimetric or not) searching for tiny spectral signatures, the detection of which can greatly benefit from the rise in S/N produced by a simultaneous extraction of the signal in many spectral lines.

To fulfill this range of ambitious technical performances, the ESPaDOnS concept was proposed by Donati et al. (1998). The instrument, developed by a collaboration including French, Canadian and Dutch experts, is now in its final assembling and testing phase at Observatoire Midi-Pyr\'en\'ees (France) before its first light scheduled at the CFHT during the second half of the year.

We first describe the main characteristics of this instrument, which includes a polarimeter mounted on the Cassegrain bonnette of the telescope, fiber linked to a bench mounted spectrograph. We then evoke some of the most promising scientific issues that such a unique facility will make it possible to address in a near future.

\section{Instrument description}

\subsection{The Cassegrain module}

In order to keep instrumental polarization to a minimum, the polarimetric module is located on the Cassegrain bonnette of the reflector, thus avoiding any oblique reflection before the polarization analysis. The Cassegrain facilities also include all calibration and guiding tools. The link to the bench-mounted spectrograph is ensured by optical fibers.

\subsubsection{The polarimeter}

Stellar light is collected at Cassegrain focus through a 1.58'' circular pinhole. Two triplets ensure that all polarizing optics work in a parallel beam and that fibers are fed at a beam aperture of f/3.6.

 The optical design of the ESPaDOnS polarimeter follows closely that of Semel's visitor stellar polarimeter used at the Anglo-Australian Telescope (Semel \& Lopez Ariste 2001). In particular, Fresnel rhomb retarders are employed to perform a very efficient polarimetric analysis over the whole spectral domain (the basic components being quarter-wave rhombs used either alone or placed side by side to constitute half-wave retarders). To enhance the rhombs achromaticity, a MgF$_2$ thin layer is deposited on one of their total reflection surfaces (King 1966). Both retardations due to the internal reflection and to the surface film are wavelength dependent, as well as the residual birefringence of the glass, so that an appropriate thickness of MgF$_2$ (of the order of 23~nm) is able to produce a resultant phase retardation equal to $90 \pm 0.5^\circ$ from 370~nm to 1,000~nm (while the phase delay of classical crystalline plates can deviate by about $20^\circ$ from the desired quarter-wave retardation on such a wide wavelength range). A dedicated procedure, adapted from the work of Williams et al. (1997), has been developed to test rhombs made from different materials and coated with different thickness of the film, in order to select the most efficient components. In addition to being the most achromatic retarders available, Fresnel rhombs do not produce detectable spectral ripples in high-resolution polarized spectra (by contrast with most  crystalline plates), thanks to a typical fringe spacing of the order of the pixel size at the detector level.

The polarization analysis is performed by a quarter-wave rhomb encompassed by two half-wave rhombs and followed by a Wollaston prism. Both half-wave rhombs can rotate about the optical axis and be oriented to predefined azimuths (regularly spaced at $22.5^\circ$ intervals) with respect to the optic axis of the Wollaston prism. The upper half-wave Fresnel rhomb can also rotate at a constant rate (at a frequency of about 1~Hz) in order to average out linear polarization when performing circular polarization analysis of highly linearly polarized sources. The two beams produced by the Wollaston prism are imaged onto the two fibers (hereafter object fibers) of the fiber link.

In non polarimetric mode, the Wollaston prism is replaced by a wedged plate with an output surface tilted to produce a beam deviation equal to that of the Wollaston prism. Note that only one of the two object fibers is used in this case (light entering the second object fiber can be blocked further on the optical path, see Section~\ref{sect:fibre}). In this mode, a second circular aperture at the Cassegrain focus (7.9" south of the central one, with a diameter of about 2'') collects photons from the sky background and redirects them to the third fiber of the fiber link (hereafter skyfibre). Sky light is also collected through this fiber in polarimetric mode, and is blocked before entering the spectrograph.

All motions of the optical components inside the polarimeter (half-wave rhombs, switch from polarimetric to non-polarimetric mode) are remote controlled. 

\subsubsection{The interface / calibration module}

An interface module is located before the polarimeter entrance to provide guiding and calibration facilities. In stellar observing mode, the beam goes through a removable atmospheric dispersion corrector if the star of interest is low on the horizon. At Cassegrain focus, photons that are not transmitted to the polarimeter are reflected off a mirror and redirected to a field viewing camera that can be used to perform a guiding either on the observed star or on another star of the field (up to a V magnitude of 19). 

In calibration mode, a removable prism is inserted on the optical axis to redirect the beam from the calibration lamps (hollow cathode and halogen lamps for wavelength and flat-field calibration) to the mirror atop the polarimeter. Polarizers, diffusers and a Fabry-Perot filter can also be inserted in the beam to provide further calibration options. The wheel containing the prism and the other calibration tools is remote controlled.

\subsection{The fiber link}
\label{sect:fibre}

The fiber link (about 30~m long), is ensured by Heraeus preform of STU type fibers, providing close to optimal transmission throughout the whole spectral domain of interest. The link contains the two object fibers and the sky fiber. Only two fibers are used simultaneously (the two object fibers in polarimetric mode, one object fiber and the sky fiber in non-polarimetric mode). A double image-slicer provides an optimal injection of both beams emerging from the fibers into the spectrograph. The light emerging from the third fiber can be masked by a remotely adjustable dekker.

Two levels of spectral resolution will be available without flux loss, by tuning the slicer to a 2 slice/fiber configuration (R = 50,000) or to a 3 slice/fiber mode (R~=~65,000). The inconvenience of the last configuration is that the spectra recorded in this mode are severely under-sampled (with 1.5 pixel for one resolved element). However, by slightly tilting the image of sliced fibers with respect to CCD lines, we ensure that the different spectral columns throughout each order are sampled on a different pixel grid. Using micro-scanning techniques as part of the extraction routines will then allow the recovery of the observed spectrum with at least twice as many sampling points, bringing the resolution element up to a comfortable size of 3.0 sub-pixels.
 
\subsection{The spectrograph}

The high resolution spectrograph of ESPaDOnS, inspired by the FEROS spectrograph now in use at La Silla Observatory (Kaufer et al. 2000), is a prism cross-dispersed \'echelle spectrograph working in dual pupil and quasi Littrow configurations, and including a fully-dioptric camera.

The advantage of the dual pupil design is to feature two different well defined polychromatic pupils, one on each dispersing device, thus producing negligible vignetting within the camera even with a clear entrance aperture only slightly larger than the pupil size. Moreover, for fiber-fed instruments such as ESPaDOnS (featuring roughly homogeneously lit pupils), using a fully dioptric camera avoids the problem of central obscuration characteristic of catadioptric cameras and further reduces the vignetting. Altogether, a very high total throughput (as much as 53\% throughout most of the spectral domain for the spectrograph alone without the CCD detector) can be obtained with such a design.

The main difference with FEROS is that the pupil size is larger (200 mm), which enables to collect stellar light through a 1.58" circular aperture with no increase in detector size nor decrease in spectral domain and resolution. The camera design is thus completely new and features 7 lenses in 4 groups, with a clear entrance aperture of 220 mm and a focal length of 388 mm. 

The main optical components of the spectrograph include a cross-disperser prism, a 79 gr/mm grating and a $2048\times4600$ 13.5 micron pixel EEV CCD detector on which the whole optical domain (370 to 1,000 nm, i.e. orders \#23 to \#61) can fit in a single exposure. A Hartmann mask is located at the entrance of the camera for precise focus adjusting, and an exposure meter takes a small fraction of the light (before dispersion) to monitor the efficiency of light injection during exposures.

The spectrograph is surrounded by a twin layer enclosure in order to ensure an optimal thermal stability throughout the nights. The first enclosure (with no air circulation) is installed on the spectrograph bench. The second layer wraps the bench itself (down to the floor) and the inner enclosure. Air circulation is implemented between the two layers. A thermal stability of order of 0.1 deg peak to valley is expected.

\subsection{Data reduction package}

A reduction package, dedicated to the reduction of both polarimetric and non-polarimetric data, has been developed for the existing generation of spectropolarimeters. This mostly automated software, called ESpRIT (Donati et al. 1997), generalizes to polarized data the optimal spectral extraction algorithms developed by Horne (1986) and Marsh (1989). This package will be available for all ESPaDOnS users, and observers will be able to perform real-time data reduction during their observing nights. 

Further data processing will also be possible during the night through the use of a cross-correlation tool (Least-Square Deconvolution, Donati et al. 1997), adapted both to polarized or unpolarized spectra. The multiplex gain in S/N provided by this technique, benefiting from the large spectral coverage of ESPaDOnS, will reach 60 for typical solar-type stars (for which thousands of spectral lines can be simultaneously recorded). Magnetic field detections, high-sensitivity radial velocity measurements and many other scientific investigations can thus be carried out during the night.

\section{Expected performances}

The global efficiency of ESPaDOnS cannot be estimated by simply considering its total throughput (including telescope and detector contributions). Several other characteristics contribute to the excellence of this instrument, such as the spectral resolution and the wavelength range (which will improve the efficiency of many observing programs based on multi-lines techniques). A way to define the effective efficiency of the instrument (and to provide relevant comparison with other spectrographs) is therefore to evaluate the product of the photon collecting surface of the telescope, the spectral resolution of the spectrograph, the spectral domain it can collect in a single exposure, and its peak throughput (at the selected spectral resolution and in nominal seeing conditions). In this context and thanks to its very wide wavelength coverage and high total throughput, ESPaDOnS will be the most efficient high resolution spectrograph available on a 4~m class telescope, and will also be competitive with several instruments on 8~m class reflectors (with global efficiency above that of HIRES at Keck and HDS at Subaru, and of the same order as UVES at the VLT). 

In particular, ESPaDOnS will be able to yield circularly and linearly polarized spectra with S/N = 100 per 3 km.s$^{-1}$ spectral bin in 1 hour for a mV = 14 star (thus offering a 5 magnitude gain with respect to the MuSiCoS spectropolarimeter available at the T\'elescope Bernard Lyot). Fast spectroscopy (with 5 minutes exposures) will be achievable at the same noise level for targets of 11$^{\rm th}$ V magnitude.

\section{Scientific issues}

\subsection{stellar magnetism}

High resolution polarized spectra are the optimal tools for studying stellar magnetic fields, since they give access both to the strength and orientation of the field. The current generation of spectropolarimeters has produced many results of prime scientific interest on stellar magnetism (see for instance the reports, in these proceedings, on several investigations conducted with MuSiCoS at the T\'elescope Bernard Lyot), and ESPaDOnS will bring adapted solutions to several limitations currently restricting this kind of prospect.

On active stars with spectral types close to that of the Sun (G to K), photospheric magnetic fields are usually distributed in an intricate pattern of regions with opposite polarities, with amplitudes of the Zeeman signatures typically below 0.3\% of the continuum level in circular polarization, even for the extremely active fast rotators (Donati et al. 2003). Multiline techniques are therefore necessary to reach the S/N required to map their photospheric field by means of Zeeman-Doppler Imaging (Donati \& Brown 1997). For a K-type star, the multiplex gain in S/N provided by ESPaDOnS will be of order of 60, which constitutes a 50\% improvement with respect to the actual generation of spectropolarimeters. This gain in efficiency (added to a high throughput, a large collecting surface and a high resolution) is opening new exciting prospects, among which the investigation of magnetic activity of young solar analogs, such as T~Tauri stars (TTS from now on). The magnetic field of TTS is believed to have a strong impact on several crucial physical processes at work in these objects, and therefore on the early stages of stellar evolution. In particular, the magnetic interaction between the star and its accretion disc is believed to deeply influence the accretion and mass loss processes occuring in TTS. Precise observations of their magnetic field will thus provide important constraints on numerical models developed to describe these mecanisms, and give original clues to probe the origin of the solar magnetic activity.

Another promising issue concerns the study of magnetic field inside cool spots of active late-type stars, by extracting Zeeman signatures from molecular lines included in ESPaDOnS spectral window. While atomic spectra give few informations about the field inside spotted regions (owing to their faint contribution to the spectra), a simultaneous inversion of atomic and molecular line masks will make it possible to obtain a complete mapping of the photospheric field, thus providing a very firm observational basis to magnetic field extrapolation models (Cameron, these proceedings).

The study of magnetism among upper main sequence stars will also benefit from the high sensitivity of ESPaDOnS. For many chemically peculiar stars (hereafter CP stars), possessing strong well-ordered magnetic fields, Zeeman signatures in linearly polarized light will be detected in single lines (while multi-lines techniques were necessary with MuSiCoS), thus opening new perspectives for modeling the fine structure of their magnetic fields. The progenitors of CP stars (probably among pre-main sequence hot stars like the relatively faint Herbig~Ae/Be stars) will be possible targets for ESPaDOnS. The sensitive investigation of this class of objects will help to probe the origin of the magnetic field of CP stars (they should in particular host strong fields, in case of a truly fossil origin of CP stars magnetism). 

\subsection{circumstellar scattering matter}

Asymmetric circumstellar material (e.g. in an accretion disk near a T Tauri star, or the circumstellar disk of a Be star) is able to polarize, both linearly and circularly, scattered light from the stellar photosphere. In particular, highly sensitive studies of how this polarization varies through emission lines will greatly help to constrain the general geometry of the scattering environment and its relationship to gaseous emission regions. 

\subsection{Indirect imaging of stellar surfaces}

Doppler imaging of stellar surfaces requires both a high spectral resolution and a low photon noise. High quality data sets are particularly needed when investigating subtle effects as photospheric differential rotation (Petit et al. 2003). Investigations with unprecedented accuracy will be carried out by ESPaDOnS in this field.

\subsection{extra-solar planets}

Spectroscopic techniques developed in the aim of detecting extra-solar planets from starlight reflection (Cameron et al. 2002) will benefit from the large spectral coverage, especially on the red side of the optical domain, to simultaneously extract the tiny planet signature from as many spectral lines as possible. A large spectral domain will also be appreciated by teams searching for planets from radial velocity fluctuations, with procedures also based on cross-correlation techniques.

\subsection{asteroseismology}

High resolution spectroscopy presents several advantages over photometry for the study of stellar oscillations. Thanks to the addition of many spectral lines, the detection of radial modes with amplitudes of only a few m s$^{-1}$ (a few cm s$^{-1}$ in Fourier space) will be achievable. Multi-lines techniques will also make it possible to monitor, with a high exposure rate, the small line profile distortions generated by high-degree modes.

ESPaDOnS will therefore be a very efficient ground-based tool for asteroseismologists, in addition to the space missions (COROT, Eddington) actually developed in Europe. Major scientific results can be expected on several issues, among which determination of internal structure and internal dynamics of solar-type stars, study of the envelopes of hot pulsating stars, or investigation of the potential connection between magnetic fields and pulsations of roAp stars (Cunha \& Gough 2000).

\section{The ESPaDOnS network}

The overall ESPaDOnS concept is expected to be duplicated on several telescopes. A first option is the copy of the polarimeter alone, fiber linked to an existing \'echelle spectrograph. A second possibility consists in an adapted copy of both polarimeter and spectrograph, an option adopted by NARVAL, already under construction at Observatoire Midi-Pyr\'en\'ees to replace MuSiCoS at TBL in 2005 (Auri\`ere, these proceedings). Other clones are currently under discussion with several institutes. 

The ESPaDOnS network will offer the opportunity to conduct multi-site observing campaigns, necessary to many studies requiring continuous observations on a timescale of a few days, such as  asteroseismology (for which continuous observing windows greatly improve the accuracy of mode detection), observing programs based on a complete covering of several stellar rotations (Doppler and Zeeman-Doppler Imaging, differential rotation measurements, search for starlight reflected by exoplanets), as well as studies of short variations of stellar objects, like monitoring of flaring events. 


\end{document}